\documentclass[pra,preprint,tightenlines,a4paper,showpacs]{revtex4}
\usepackage{bm,dcolumn,graphicx}
\begin{document}
%####################################################################
\title{High-precision calculations of electric-dipole amplitudes
for transitions between low-lying levels of Mg, Ca, and Sr}
\author{S.~G.~Porsev}
\email{porsev@thd.pnpi.spb.ru},
\author{M.~G.~Kozlov}
\author{Yu.~G.~Rakhlina}
\affiliation{Petersburg Nuclear Physics Institute,
Gatchina, Leningrad district, 188300, Russia}
\author{A.~Derevianko}
\affiliation{Physics Department, University of Nevada, Reno,
Nevada 89557}
\date{\today}

\begin{abstract}
To support efforts on  cooling and trapping of alkaline-earth
atoms and designs of atomic clocks,  we performed \emph{ab initio}
relativistic many-body calculations of electric-dipole transition
amplitudes between low-lying states of Mg, Ca, and Sr. In
particular, we report amplitudes for $^1P^o_1 \rightarrow
{}^1S_0,{}^3S_1,{}^1D_2$, for $^3P^o_1 \rightarrow
{}^1S_0,{}^1D_2$, and for $^3P^o_2 \rightarrow {}^1D_2$
transitions. For Ca, the reduced matrix element  $\langle 4s4p
\,^1\! P^o_1||D||4s^2 \,^1\! S_0 \rangle$ is in a good agreement
with a high-precision experimental value deduced from
photoassociation spectroscopy [Zinner {\em et al.}, Phys.\ Rev.\
Lett.\ {\bf 85}, 2292 (2000) ]. An estimated uncertainty of the
calculated lifetime of the $3s3p\,^1\! P^o_1$ state of Mg is a
factor of three smaller than that of the most accurate experiment.
Calculated binding energies reproduce experimental values within
0.1-0.2\%.
\end{abstract}


\maketitle
%####################################################################

%=====================
\section{Introduction}
%=====================
Many-body methods have proven to be a highly accurate tool for
determination of atomic properties, especially for systems with
one valence electron outside a closed-shell core~\cite{Sap98}. For
alkali-metal atoms a comparison of highly-accurate experimental
data with calculations~\cite{SafJohDer99} allows one to draw a
conclusion that modern \emph{ab initio} methods are capable of
predicting basic properties of low-lying states with a precision
better than 1\%.

For {\em divalent} atoms  such a comprehensive comparison was
previously hindered by a lack of high-precision measurements of
radiative lifetimes. Despite the lifetimes of the lowest
$nsnp\,^1\!P^o_1$ and $nsnp\,^3\!P^o_1$ states were repeatedly
obtained both experimentally and
theoretically \cite{Zin,Seng,L80,FF,JonFisGod99,VaeGodHan88,KM80,BFV,WGTG,
Lund,PRT,Smith,Hans, Hunter,God,HusR,HusS,Kwong,Droz,Mitch,Whit,
Kell}, persistent discrepancies remain. Only very recently, Zinner
{\em et al.}~\cite{Zin} have achieved 0.4\% accuracy for the rate
of $4s4p\,^1\!P^o_1 \rightarrow 4s^2\,^1\!S_0$ transition in
calcium. This high-precision value was deduced from
photoassociation spectroscopy of ultracold calcium atoms. One of
the purposes of the present work is to test the quality of
many-body techniques for two-valence electron systems by comparing
our result with the experimental value from Ref.~\cite{Zin}.

We extend the earlier work~\cite{PorKozRah00} and report results
of  relativistic many-body calculation of energy levels and
electric-dipole transition amplitudes for Mg, Ca and Sr. The
calculations are performed in the framework of
configuration-interaction approach coupled with many-body
perturbation theory~\cite{DFK,DKPF}. We tabulate
electric-dipole amplitudes for $^1P^o_1 \rightarrow
{}^1S_0,{}^3S_1,{}^1D_2$, for $^3P^o_1 \rightarrow
{}^1S_0,{}^1D_2$, and for $^3P^o_2 \rightarrow {}^1D_2$
transitions and estimate theoretical uncertainties.

Cooling and trapping experiments with alkaline-earth atoms were
recently reported for Mg~\cite{Seng}, Ca~\cite{Zin,Kuro}, and
Sr~\cite{Din,Kat}. The prospects of achieving Bose-Einstein
condensation were also discussed~\cite{Mg,Zin}. Our accurate
transition amplitudes will be helpful in designs of cooling
schemes and atomic clocks. In addition, these amplitudes will aid
in determination of long-range atomic interactions, required in
calculation of scattering lengths and interpretation of
cold-collision data. For example, dispersion (van der Waals)
coefficient $C_6$ characterizes the leading dipole-dipole
interaction of two ground-state atoms at large internuclear
separations~\cite{DalDav66}.  The coefficient $C_6$ is expressed
in terms of energy separations and electric-dipole matrix elements
between the ground and excited atomic states. Approximately 80\%
of the total value of $C_6$ arises from the principal transition
$nsnp\,^1\!P^o_1 - ns^2\,^1\!S_0$, requiring accurate predictions
for the relevant matrix element. Therefore our results  will be
also useful in determination of dispersion coefficients.

%===============================
\section{Method of calculations}
%===============================

In atomic-structure calculations, correlations are conventionally
separated into three classes: valence-valence, core-valence, and
core-core correlations. A strong repulsion of valence electrons
has to be treated non-perturbatively, while it is impractical to
handle the two other classes of correlations with non-perturbative
techniques, such as configuration-interaction (CI) method.
Therefore, it is natural to combine many-body perturbation theory
(MBPT) with one of the non-perturbative methods. It was
suggested \cite{DFK} to use MBPT to construct an effective
Hamiltonian $H_{\rm eff}$ defined in the model space of valence
electrons. Energies and wavefunctions of low-lying states are
subsequently determined using CI approach, i.e. diagonalizing
$H_{\rm eff}$ in the valence subspace. Atomic observables are
calculated with effective operators~\cite{DKPF}. Following the
earlier work, we refer to this method as CI+MBPT formalism.

In the CI+MBPT approach the energies and wavefunctions are
determined from the Schr\"odinger equation
\begin{equation} \label{Eqn_Sh}
H_{\rm eff}(E_n) \, | \Phi_n \rangle = E_n \, |\Phi_n \rangle \, ,
\end{equation}
where the effective Hamiltonian is defined as
\begin{equation}\label{Eqn_Heff}
  H_{\rm eff}(E) = H_{\rm FC} + \Sigma(E).
\end{equation}
Here $H_{\rm FC}$ is the two-electron Hamiltonian in the frozen
core approximation and $\Sigma$ is the
energy-dependent correction, involving core excitations. The
operator $\Sigma$ completely accounts for the second order of
perturbation theory. Determination of the second order corrections
requires calculation of one-- and two--electron diagrams. The
one--electron diagrams describe an attraction of a valence
electron by a (self-)induced core polarization. The two-electron
diagrams are specific for atoms with several valence electrons and
represent an interaction of a valence electron with core
polarization induced by another valence electron.

Already at the second order the number of the two--electron
diagrams is large and their computation is very time-consuming. In
the higher orders the calculation of two-electron diagrams becomes
impractical. Therefore we account for the higher orders of MBPT
indirectly. It was demonstrated~\cite{Opt}  that a proper
approximation for the effective Hamiltonian can substantially
improve an agreement between calculated and experimental spectra
of multielectron atom. One can introduce an energy shift $\delta$
and replace $\Sigma(E) \rightarrow \Sigma(E-\delta)$ in the
effective Hamiltonian, Eq.~(\ref{Eqn_Heff}).
The choice $\delta$=0 corresponds to the Brillouin-Wigner variant
of MBPT  and the Rayleigh-Schr\"odinger variant is recovered
setting $\delta = E_n - E_n^{(0)}$, where $E_n^{(0)}$ is the
zero-order energy of level $n$. The latter is more adequate for
multielectron systems~\cite{Thoul}; for few-electron systems an
intermediate value of $\delta$ is optimal. We have determined
$\delta$ from a fit of theoretical energy levels to experimental
spectrum. Such an optimized effective Hamiltonian was used in
calculations of transition amplitudes.

To obtain an effective electric-dipole operator we solved
random-phase approximation (RPA) equations, thus summing a certain
sequence of many-body diagrams to all orders of MBPT. The RPA
describes a shielding of externally applied field by core
electrons. We further incorporated one- and two-electron
corrections to the RPA to account for a difference between the
V$^N$ and V$^{N-2}$ potentials and for the Pauli exclusion
principle. In addition, the effective operator included
corrections for normalization  and structural
radiation~\cite{DKPF}. The RPA equations depend on transition
frequency and should be solved independently for each transition.
However, the frequency dependence was found to be rather weak and
we solved these equations only at some characteristic frequencies.
To monitor a consistency of the calculations we employed both
length (L) and velocity (V) gauges for the electric-dipole operator.

The computational procedure is similar to calculations of
hyperfine structure constants and electric-dipole amplitudes for
atomic ytterbium \cite{PRK1,PRK2}. We consider Mg, Ca and Sr as
atoms with two valence electrons above closed cores
[1$s$,...,2$p^6$], [1$s$,...,3$p^6$], and [1$s$,...,4$p^6$],
respectively \cite{sidenote}.
One-electron basis set for Mg included 1$s$--13$s$, 2$p$--13$p$,
3$d$--12$d$, and 4$f$--11$f$ orbitals, where the core- and 3,4$s$,
3,4$p$, 3,4$d$, and 4$f$ orbitals were Dirac-Hartree-Fock (DHF)
ones, while all the rest were virtual orbitals. The orbitals
1$s$--3$s$ were constructed by solving the DHF equations in V$^N$
approximation, 3$p$ orbitals were obtained in the V$^{N-1}$
approximation, and 4$s$, 4$p$, 3,4$d$, and 4$f$ orbitals were
constructed in the V$^{N-2}$ approximation. We determined virtual
orbitals using a recurrent procedure, similar to
Ref.~\cite{Bogdan} and employed in previous
work~\cite{DFK,DKPF,PRK1,PRK2}. The one-electron basis set for Ca
included 1$s$--13$s$, 2$p$--13$p$, 3$d$--12$d$, and 4$f$--11$f$
orbitals, where the core- and 4$s$, 4$p$, and 3$d$ orbitals are
DHF ones, while the remaining orbitals are the virtual orbitals.
The orbitals 1$s$--4$s$ were constructed by solving the DHF
equations in the V$^N$ approximation, and 4$p$ and 3$d$ orbitals
were obtained in the V$^{N-1}$ approximation. Finally, the
one-electron basis set for Sr included 1$s$--14$s$, 2$p$--14$p$,
3$d$--13$d$, and 4$f$--13$f$ orbitals, where the core- and 5$s$,
5$p$, and 4$d$ orbitals are DHF ones, and all the rest are the
virtual orbitals. The orbitals 1$s$--5$s$ were constructed by
solving the DHF equations in the V$^N$ approximation, and 5$p$ and
4$d$ orbitals were obtained in the V$^{N-1}$ approximation.
Configuration-interaction states were formed using these
one-particle basis sets. It is worth emphasizing that the employed
basis sets were sufficiently large to obtain numerically converged
CI results. A numerical solution of random-phase approximation
equations required an increase in the number of virtual
orbitals. Such extended basis sets included $1s$--$ks$,
$2p$--$kp$, $3d$--($k$-1)$d$, $4f$--($k$-4)$f$, and
$5g$--($k$-8)$g$ orbitals, where $k$=19,20,21 for Mg, Ca, and Sr,
respectively. Excitations from all core shells were included in
the RPA setup.

%===============================
\section{Results and discussion}
%===============================
\subsection{Energy levels}
In Tables~\ref{Mg_E} -- \ref{Sr_E} we present calculated energies
of low-lying states for Mg, Ca, and Sr and compare them with
experimental values. The two-electron binding energies were
obtained both in the framework of conventional
configuration-interaction method and using the formalism of
CI coupled with many-body perturbation theory. Already
at the CI stage the agreement of the calculated and experimental
energies is at the level of 5\%. The addition of many-body
corrections to the Hamiltonian improves the accuracy by
approximately an order of magnitude. Finally,  with an optimal
choice of parameter $\delta$ the agreement with experimental
values improves to 0.1--0.2\%.

Compared to the binding energies, fine-structure splitting of
triplet states and singlet-triplet energy differences represent a
more stringent test of our method. For the $^3P^o_{1,2,3}$-states
the fine-structure splitting is reproduced with an accuracy of
several per cent in the pure CI for all the three atoms, while the
$^3P^o_1$ -- $^1P^o_1$ energy differences are less accurate
(especially for Ca and Sr). As demonstrated in Ref.~\cite{Opt}, the
two-electron exchange Coulomb integral $R_{np,ns,ns,np}$
($n$=3,4,5 for Mg, Ca, and Sr, respectively) determining the
splitting between $^3P^o_1$ and $^1P^o_1$ states is very sensitive
to many-body corrections. Indeed, with these corrections included,
the agreement with the experimental data improves to 1-2\% for all
the three atoms.

The case of the even-parity $^{3,1}D_J$-states is even more
challenging. For Ca, these four states are practically degenerate
at the CI stage. A repulsion of the level $^1D_2$ from the
upper-lying levels of  $np^2$ configuration pushes it down to the
level $^3D_2$ and causes their strong mixing. As seen from
Table~\ref{Ca_E} these states are separated only by 10 cm$^{-1}$,
while the experimental energy difference  is 1550 cm$^{-1}$. As a
result, an accurate description of superposition of $^3D_2$ and
$^1D_2$ states is important. The $^3D_2$ -- $^1D_2$ splitting is
restored when the many-body corrections are included in the
effective Hamiltonian. These corrections properly account for core
polarization screening an interaction between $sd$ and $p^2$
configurations.

For Sr, the fine-structure splittings of $^3D_J$ states and energy
difference between the $^3D_J$ and the $^1D_2$ levels are also
strongly underestimated in the pure CI method. Again the inclusion
of the many-body corrections substantially improves the splittings
between the $D$-states. It is worth emphasizing, that for such an
accurate analysis a number of effects was taken into account,
i.e., spin-orbit interaction, configuration interaction, and
core-valence correlations. A proper account for all these effects
is of particular importance for determination of electric-dipole
amplitudes forbidden in $LS$-coupling, such as for $^3P^o_J
\rightarrow {}^1S_0,{}^1D_2$ transitions.

\subsection{Transition amplitudes}

In this section we present calculations of electric-dipole (E1)
amplitudes for $^{3,1}P^o_1 \rightarrow {}^1S_0$, $^{3,1}P^o_1
\rightarrow {}^1D_2$, $^3P^o_2 \rightarrow {}^1D_2$, and $^1P^o_1
\rightarrow {}^3S_1$ transitions. The calculated reduced matrix
elements for Mg, Ca, and Sr are presented in
Tables~\ref{Tab_E1_allowed} and \ref{Tab_E1_inter}. For a
transition $I \rightarrow F$ the Einstein rate coefficients for
spontaneous emission (in $1/s$) are expressed in terms of these
reduced matrix elements $\langle F || D || I \rangle$ (a.u.) and
wavelengths $\lambda$ (\AA) as
\begin{equation}
A_{FI} = \frac{2.02613 \times 10^{18}} {\lambda^3 }
 \frac{ |\langle F || D || I \rangle|^2}{ 2 J_I +1} \, .
\end{equation}
A number of long-range atom-atom interaction coefficients  could
be directly obtained from the calculated matrix elements. At large
internuclear separations $R$ an atom in a state $|A \rangle$
predominantly interacts with a like atom in a state $|B\rangle$
through a potential $V(R) \approx \pm C_3/R^3$, provided an
electric-dipole transition between the two atomic states $|A
\rangle$ and $|B \rangle$ is allowed. The coefficient $C_3$ is
given by
\begin{equation}
|C_3| = |\langle A||D||B\rangle|^{2} \, \sum_{\mu=-1}%
^{1}\left(  1+\delta_{\mu,0}\right)  \left(
\begin{tabular}
[c]{ccc}%
$J_{A}$ & $1$ & $J_{B}$\\
$-\frac{\Omega+\mu}{2}$ & $\mu$ & $\frac{\Omega-\mu}{2}$%
\end{tabular}
\right)  ^{2}%
\, ,
\end{equation}
where $\Omega$ is the conventionally defined sum of projections of
total angular momenta on internuclear axis.

From a solution of the eigen-value problem, Eq.~(\ref{Eqn_Sh}), we
obtained wave functions, constructed effective dipole operators,
and determined the transition amplitudes. The calculations were
performed within both traditional configuration-interaction
method and CI coupled with the many-body perturbation theory.
The comparison of the CI and the CI+MBPT values allows
us to estimate an accuracy of our calculations. As it was
mentioned above, to monitor the consistency of the calculations, we
determined the amplitudes using both length and velocity gauges
for the dipole operator. In general, dipole amplitudes calculated in
the velocity gauge are more sensitive to many-body corrections; we
employ the length form of the dipole operator in our final tabulation.

We start the discussion with the amplitudes for the principal
$nsnp\,^1\!P^o_1 \rightarrow ns^2\, ^1\!S_0$ transitions ($n=3$
for Mg, $n=4$ for Ca, and $n=5$ for Sr). Examination of
Table~\ref{Tab_E1_allowed} reveals that the many-body effects
reduce the L-gauge amplitudes by 1.6\% for Mg, 5.5\% for Ca,
and 6.4\% for Sr. Further, the MBPT corrections bring the length
and velocity-form results into a closer agreement. For example,
for Sr at the CI level the velocity and length forms differ by
2.7\% and this discrepancy is reduced to 0.8\% in the CI+MBPT
calculations.

A dominant theoretical uncertainty of the employed CI+MBPT method
is due to impossibility to account for all the orders of many-body
perturbation theory. It is worth emphasizing that in our CI
calculations the basis sets were saturated and the associated
numerical errors were  negligible. We expect that the theoretical
uncertainty is proportional to the determined many-body
correction. In addition, we take into account the proximity of the
amplitudes obtained in the L- and V-gauges.  We estimate the
uncertainties for the $nsnp\,^1\!P^o_1 \rightarrow ns^2\, ^1\!S_0$
transition amplitudes as 25--30\% of the many-body corrections in
the length gauge. The final values for $\langle nsnp\,^1\!P^o_1
||D|| ns^2\, ^1\!S_0 \rangle$, recommended from the present work,
are 4.03(2) for Mg, 4.91(7) for Ca, and 5.28(9) a.u. for Sr.

\begin{figure}
\centerline{\includegraphics*[scale=0.75]{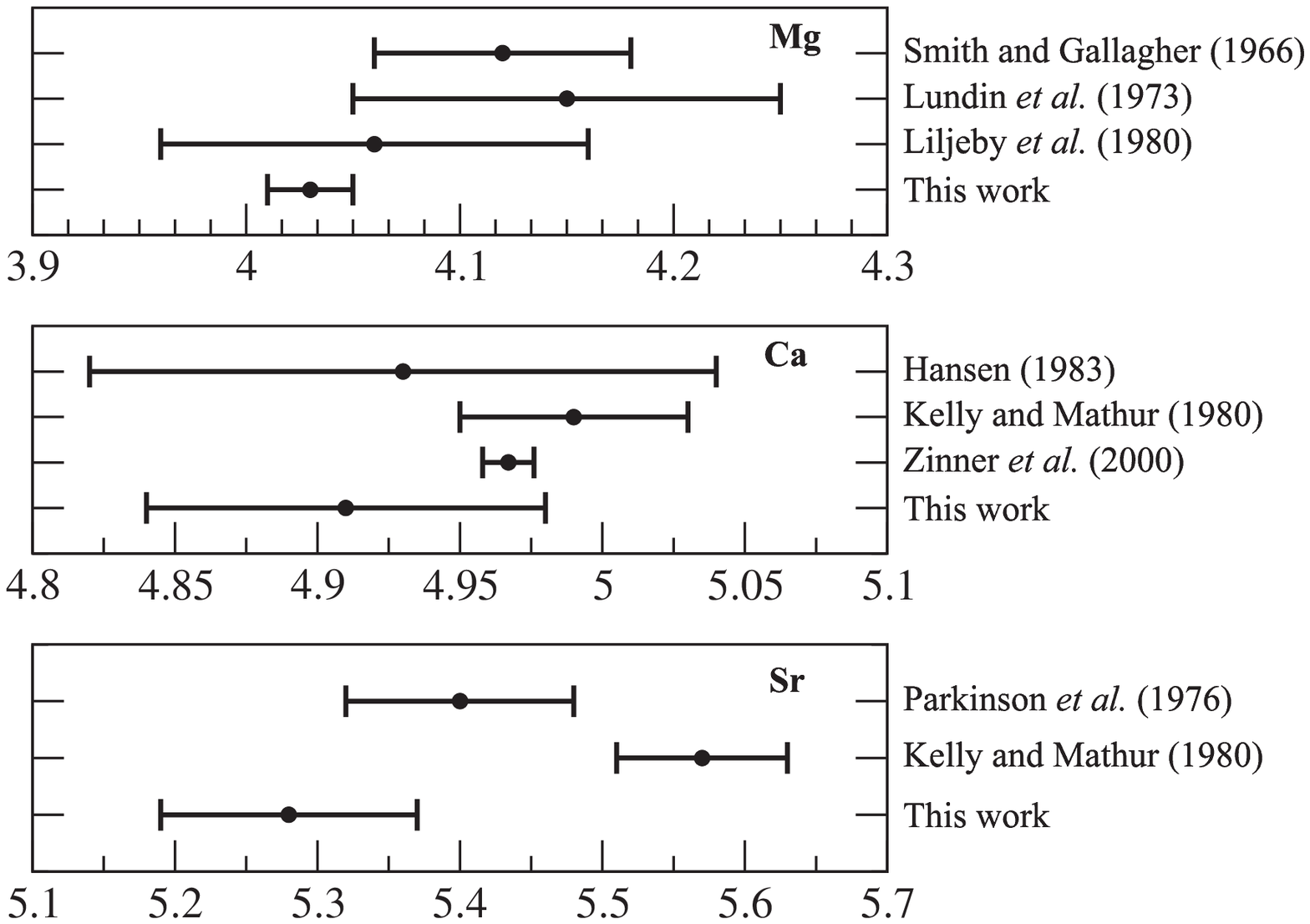}}
\caption{ Comparison of calculated reduced matrix elements
$\langle nsnp\,^1\!P^o_1 ||D|| ns^2\, ^1\!S_0 \rangle$
with experimental data in a.u.
\label{Fig_princ} }
\end{figure}

We present a comparison of our results for $\langle
nsnp\,^1\!P^o_1 ||D|| ns^2\, ^1\!S_0 \rangle$  with experimental
data in Table~\ref{Tab_E1_allowed} and in Fig.~\ref{Fig_princ}.
Our estimated accuracy for Mg is a factor of three better than
that of the most accurate experiment and for Sr is comparable to
experimental precision. For Ca, the dipole matrix element of the
$^1\!P^o_1 \rightarrow {}^1\!S_0$ was recently determined with a
precision of 0.2\% by Zinner {\em et al.} \cite{Zin} using
photoassociation spectroscopy of ultracold Ca atoms.  While our
result is in harmony with their value, the experimental accuracy
is substantially better. An updated analysis~\cite{Tiemann} of
photoassociation spectra of Ref.~\cite{Zin} leads to a somewhat
better agreement with our calculated value.

A very extensive compilation of earlier theoretical results for
the $ ^1\!P^o_1 \rightarrow {}^1\!S_0$ transition amplitudes can be
found in Ref.~\cite{FF} for Mg and in Ref.~\cite{BFV} for Ca. In a
very recent multiconfiguration Hartree-Fock (MCHF) calculations
for Mg~\cite{JonFisGod99} the authors have determined $ \langle
3s3p\,^1\!P^o_1||D|| 3s^2\,^1\!S_0 \rangle =4.008 $ a.u. This value
agrees with our final result of 4.03(2) a.u. For heavier Sr the
correlation effects are especially pronounced and only a few
calculations were performed. For example, MCHF calculations for
Sr~\cite{VaeGodHan88} found in the length gauge $ \langle
5s5p\,^1\!P^o_1||D|| 5s^2\,^1\!S_0 \rangle = 5.67$ a.u. By contrast
to the present work, the core-polarization effects were not
included in this analysis. As a result, this calculated value is
in a good agreement with our result 5.63 a.u. obtained at the CI stage,
but differs from the final value $5.28(9)$ a.u.

Another nonrelativistically allowed transition is $^1P^o_1
\rightarrow {}^1D_2$ and one could expect that this amplitude can
be determined with a good accuracy. For Mg this is really so.
However, for Ca and Sr an admixture of the configuration $p^2$
brings about large corrections to this amplitude, especially in
the velocity gauge. Another complication is the following. The
matrix element of electric-dipole operator can be represented in
the V-gauge as (atomic units $\hbar = |e| = m_e = 1$ are used):
\begin{equation}
\langle F |{\bf D}| I \rangle = i \,c \, \langle F|{\bm{\alpha}}|
I \rangle /(E_I - E_F).
\end{equation}
Here $c$ is the speed of light, $E_I$ and $E_F$ are the energies
of initial and final states, and $\bm{\alpha}$ are the Dirac
matrices. For the $^1P^o_1 \rightarrow {}^1D_2$ transition in Ca
and Sr the energy denominator is approximately 0.01 a.u.  Because
the E1-amplitudes of these transitions $\sim 1$ a.u. (see
Table~\ref{Tab_E1_allowed}), the respective numerators are of the
order of 0.01 a.u. Correspondingly the  matrix elements $\langle
F|{\bm{\alpha}}|I\rangle$ are small and are very sensitive to
corrections, i.e., the V-gauge results are unstable. As a result
we present only the L-gauge values for $^1P^o_1 \rightarrow
{}^1D_2$ E1 amplitudes for Ca and Sr. An absence of reliable
results in V-gauge hampers an estimate of the accuracy, so we
rather conservatively take it to be 25\%. Note that even with such
a large uncertainty our value for Sr significantly differs from
the experimental value \cite{Hunter}. The measurement in
\cite{Hunter} has been carried out on the $^1D_2 \rightarrow
{}^1S_0$ transition and an interference between
electric-quadrupole (E2) and Stark-induced dipole amplitudes was
observed. In order to determine the transition rate a theoretical
value of the E2-amplitude for the $^1D_2 \rightarrow {}^1S_0$
transition was taken from \cite{Bausch}. It may be beneficial
either to measure directly the rate of the E1-transition $^1P^o_1
\rightarrow {}^1D_2$ or to measure the rate of the E2-transition
$^1D_2 \rightarrow {}^1S_0$.

For the $^3P^o_J \rightarrow {}^1S_0,{}^1D_2$ transitions the
respective E1-amplitudes are small; these are nonrelativistically
forbidden intercombination transitions and consequently their
amplitudes are proportional to spin-orbit interaction. The
calculated reduced matrix elements are presented in
Table~\ref{Tab_E1_inter}.

One can see from Tables~\ref{Mg_E} --~\ref{Sr_E} that the MBPT
corrections to the fine structure splittings are large, amplifying
significance of higher order many-body corrections. In addition,
higher order corrections in the fine-structure constant $\alpha$
to the Dirac-Coulomb Hamiltonian are also important here. As
demonstrated in Ref.~\cite{FF}, the Breit interaction reduces the
dipole amplitude of $^3P^o_1 \rightarrow {}^1S_0$ transition in Mg
by 5\%. At the same time for all the intercombination transitions
the agreement between L- and V-gauges is at the level of 6-8\%. We
conservatively estimate the uncertainties of the calculated
intercombination E1 amplitudes to be 10--12\%.

To reiterate, we carried out calculations of energies of low-lying
levels and electric-dipole amplitudes between them for divalent
atoms Mg, Ca,  and Sr. We employed {\em ab initio} relativistic
configuration interaction method coupled with many-body
perturbation theory. The calculated removal energies reproduce
experimental values within 0.1-0.2\%. A special emphasis has been
put on accurate determination of electric-dipole amplitudes for
principal transitions $nsnp\,^1\!P^o_1 \rightarrow ns^2\,^1\!S_0$.
For these transitions, we estimated theoretical uncertainty to be
0.5\% for Mg, 1.4\% for Ca, and 1.7\% for Sr. For Ca, the reduced
matrix element  $\langle 4s4p \,^1\! P^o_1||D||4s^2 \,^1\! S_0
\rangle$ is in a good agreement with a high-precision experimental
value~\cite{Zin}. An estimated uncertainty of the calculated
lifetime of the lowest $^1\! P^o_1$ state for Mg is a factor of
three smaller than that of the most accurate experiment. In
addition, we evaluated electric-dipole amplitudes and estimated
theoretical uncertainties for $^1P^o_1 \rightarrow
{}^3\!S_1,{}^1\!D_2$, $^3\!P^o_1 \rightarrow {}^1\!S_0,{}^1D_2$,
and for $^3\!P^o_2 \rightarrow {}^1\!D_2$ transitions. Our results
could be useful in designs of cooling schemes and atomic clocks,
and for accurate description of long-range atom-atom interactions
needed in interpretation of cold-collision data.

\acknowledgments We would like to thank H. Katori, C. Oates, and
F. Riehle for stimulating discussions. This work was supported in
part by the Russian Foundation for Basic Researches (grant No
98-02-17663). The work of A.D. was partially supported by the
Chemical Sciences, Geosciences and Biosciences Division of the
Office of Basic Energy Sciences, Office of Science, U.S.
Department of Energy.

%####################################################################
\begin{table}
\caption{Two-electron binding energies E$_{\rm val}$ in a.u. and
energy differences $\Delta$ (cm$^{-1}$) for low-lying levels of
Mg.}
\label{Mg_E}
%\begin{ruledtabular}
\begin{tabular}{lcdcdcdc}
\hline \hline
& & \multicolumn{2}{ c}{\qquad CI}
& \multicolumn{2}{ c}{\qquad CI+MBPT}
& \multicolumn{2}{c}{\qquad Experiment} \\
Config.& Level
&\multicolumn{1}{c}{\qquad E$_{\rm val}$}
&\multicolumn{1}{c}{$\Delta$}
&\multicolumn{1}{c}{\qquad E$_{\rm val}$}
&\multicolumn{1}{c}{$\Delta$}
&\multicolumn{1}{c}{\qquad E$_{\rm val}$}
&\multicolumn{1}{c}{$\Delta$}  \\
\hline
$3s^2$     &  $^1S_0$  &  0.819907  &   ---
& 0.833556 &    ---    &  0.833518 \footnotemark[1] & --- \\
$3s\,4s$   &  $^3S_1$  &  0.635351  &  40505
& 0.645853 &   41196   &  0.645809  &  41197.4 \\
$3s\,4s$   &  $^1S_0$  &  0.624990  &  42779
& 0.635283 &   43516   &  0.635303  &  43503.1 \\
$3s\,3d$   &  $^1D_2$  &  0.613603  &  45278
& 0.621830 &   46469   &  0.622090  &  46403.1 \\
\hline
$3s\,3p$   &  $^3P^o_0$ &  0.724170  &  21012
& 0.733896 &   21879    &  0.733961  &  21850.4 \\
$3s\,3p$   &  $^3P^o_1$ &  0.724077  &  21032
& 0.733796 &   21901    &  0.733869  &  21870.4 \\
$3s\,3p$   &  $^3P^o_2$ &  0.723889  &  21073
& 0.733596 &   21945    &  0.733684  &  21911.1 \\
$3s\,3p$   &  $^1P^o_1$ &  0.662255  &  34601
& 0.674226 &   34975    &  0.673813  &  35051.4 \\
\hline \hline
\end{tabular}
%\end{ruledtabular}
\footnotemark[1]{Two~electron binding energy of the ground
state is determined as a sum of the first two ionization
potentials IP~(Mg) and IP~(Mg$^+$), where IP~(Mg) = 61669.1
cm$^{-1}$ and IP~(Mg$^+$)$ = 121267.4$ cm$^{-1}$ \cite{Moore}}.
\end{table}
%####################################################################
\begin{table}
\caption{Two-electron binding energies in a.u. and energy
differences $\Delta$  in cm$^{-1}$ for the low-lying levels of
Ca.}
\label{Ca_E}
%\begin{ruledtabular}
\begin{tabular}{lcdcdcdc}
\hline \hline
& & \multicolumn{2}{ c}{\qquad CI \footnotemark[1]}
& \multicolumn{2}{ c}{\qquad CI+MBPT}
& \multicolumn{2}{c}{\qquad Experiment} \\
Config.& Level
&\multicolumn{1}{c}{\qquad E$_{\rm val}$}
&\multicolumn{1}{c}{$\Delta$}
&\multicolumn{1}{c}{\qquad E$_{\rm val}$}
&\multicolumn{1}{c}{$\Delta$}
&\multicolumn{1}{c}{\qquad E$_{\rm val}$}
&\multicolumn{1}{c}{$\Delta$}  \\
\hline
$4s^2$     &  $^1S_0$  &  0.636590  &   ---
& 0.661274 &    ---    &  0.660927 \footnotemark[2] & --- \\
$4s\,3d$   &  $^3D_1$  &  0.528838  &  23649
& 0.567744 &   20527   &  0.568273  &  20335.3 \\
$4s\,3d$   &  $^3D_2$  &  0.528868  &  23642
& 0.567656 &   20547   &  0.568209  &  20349.2 \\
$4s\,3d$   &  $^3D_3$  &  0.528820  &  23653
& 0.567517 &   20577   &  0.568110  &  20371.0 \\
$4s\,3d$   &  $^1D_2$  &  0.528824  &  23652
& 0.559734 &   22285   &  0.561373  &  21849.6 \\
$4s\,5s$   &  $^3S_1$  &  0.498205  &  30372
& 0.517490 &   31557   &  0.517223  &  31539.5 \\
\hline
$4s\,4p$   &  $^3P^o_0$ &  0.574168  &  13700
& 0.591521 &   15309    &  0.591863  &  15157.9 \\
$4s\,4p$   &  $^3P^o_1$ &  0.573942  &  13750
& 0.591274 &   15363    &  0.591625  &  15210.1 \\
$4s\,4p$   &  $^3P^o_2$ &  0.573486  &  13850
& 0.590774 &   15473    &  0.591143  &  15315.9 \\
$4s\,4p$   &  $^1P^o_1$ &  0.530834  &  23211
& 0.553498 &   23654    &  0.553159  &  23652.3 \\
\hline \hline
\end{tabular}
%\end{ruledtabular}
\small \footnotemark[1]{Note~that~the conventional CI fails to
recover the correct ordering of $D$-states.}
\footnotemark[2]{For the ground state E$_{\rm val}$=
IP~(Ca)+IP~(Ca$^+$), where IP~(Ca) = 49304.8 cm$^{-1}$ and IP~(Ca$^+$)
= 95752.2 cm$^{-1}$ \cite{Moore}}.
\end{table}
%####################################################################
\begin{table}
\caption{Two-electron binding energies in a.u. and energy
differences $\Delta$ in cm$^{-1}$ for the low-lying levels of Sr.}
\label{Sr_E}
%\begin{ruledtabular}
\begin{tabular}{lcdcdcdc}
\hline \hline
& & \multicolumn{2}{ c}{\qquad CI}
& \multicolumn{2}{ c}{\qquad CI+MBPT}
& \multicolumn{2}{c}{\qquad Experiment} \\
Config.& Level
&\multicolumn{1}{c}{\qquad E$_{\rm val}$}
&\multicolumn{1}{c}{$\Delta$}
&\multicolumn{1}{c}{\qquad E$_{\rm val}$}
&\multicolumn{1}{c}{$\Delta$}
&\multicolumn{1}{c}{\qquad E$_{\rm val}$}
&\multicolumn{1}{c}{$\Delta$}  \\
\hline
$5s^2$     &  $^1S_0$   & 0.586538   &   ---
& 0.614409 &    ---     &  0.614601 \footnotemark[1] & --- \\
$5s\,4d$   &  $^3D_1$   &  0.497148  &  19619
& 0.532110 &   18063    &  0.531862  &  18159.1 \\
$5s\,4d$   &  $^3D_2$   &  0.497077  &  19635
& 0.531809 &   18129    &  0.531590  & 18218.8 \\
$5s\,4d$   &  $^3D_3$   &  0.496941  &  19664
& 0.531298 &   18242    &  0.531132  & 18319.3 \\
$5s\,4d$   &  $^1D_2$   &  0.494339  &  20235
& 0.522311 &   20213    &  0.522792  & 20149.7 \\
$5s\,6s$   &  $^3S_1$   &  0.460940  &  27566
& 0.481533 &   29162    &  0.482291  & 29038.8 \\
\hline
$5s\,5p$   &  $^3P^o_0$  &  0.529636  &  12489
& 0.548754 &    14410    &  0.549366  & 14317.5 \\
$5s\,5p$   &  $^3P^o_1$  &  0.528850  &  12662
& 0.547896 &    14598    &  0.548514  & 14504.4 \\
$5s\,5p$   &  $^3P^o_2$  &  0.527213  &  13021
& 0.546079 &    14997    &  0.546718  & 14898.6 \\
$5s\,5p$   &  $^1P^o_1$  &  0.491616  &  20833
& 0.515901 &    21621    &  0.515736  & 21698.5 \\
\hline \hline
\end{tabular}
%\end{ruledtabular}
\small \footnotemark[1]{For~the~ground state E$_{\rm val}$ =
IP~(Sr)+IP~(Sr$^+$), where IP~(Sr) = 45925.6 cm$^{-1}$ and
IP~(Sr$^+$) = 88964.0 cm$^{-1}$ \cite{Moore}}.
\end{table}

%####################################################################
\begin{table}
\caption{ Reduced electric-dipole matrix elements for transitions
allowed in $LS$-coupling. $n$ is the principal quantum number of
the first valence $s$ and $p$ shells and $m$ corresponds to the
first valence $d$ shell; $n=3$ for Mg, 4 for Ca,  and 5 for Sr;
$m=3$ for Mg and Ca, and 4 for Sr. All values are in a.u.}
\label{Tab_E1_allowed}
%\begin{ruledtabular}
\begin{tabular}{ldddddd}
\hline \hline
&\multicolumn{2}{c}{Mg} & \multicolumn{2}{c}{Ca} &
\multicolumn{2}{c}{Sr}
\\
&\multicolumn{1}{c}{CI} & \multicolumn{1}{c}{CI+MBPT} &
\multicolumn{1}{c}{CI} & \multicolumn{1}{c}{CI+MBPT} &
\multicolumn{1}{c}{CI} & \multicolumn{1}{c}{CI+MBPT}
\\
\hline \multicolumn{7}{c}{$\langle nsnp
\,^1\!P^o_1 ||D|| ns^2\, ^1\!S_0 \rangle$}\\
  L-gauge      & 4.09  & 4.03 & 5.20 & 4.91 & 5.63 & 5.28 \\
  V-gauge      & 4.06  & 4.04 & 5.11 & 4.93 & 5.48 & 5.32 \\
 Final value       & \multicolumn{2}{d }{4.03(2)}
                    & \multicolumn{2}{d }{4.91(7)}
                    & \multicolumn{2}{d}{5.28(9)} \\
 Experiment    & \multicolumn{2}{d }{4.15(10) \footnotemark[1]}
                & \multicolumn{2}{d }{4.967(9) \footnotemark[2]}
                & \multicolumn{2}{d }{5.57(6)  \footnotemark[3]}
\\             & \multicolumn{2}{d }{4.06(10) \footnotemark[4]}
                & \multicolumn{2}{d }{4.99(4)  \footnotemark[3]}
                & \multicolumn{2}{d }{5.40(8)  \footnotemark[5]}
\\             & \multicolumn{2}{d }{4.12(6)  \footnotemark[6]}
                & \multicolumn{2}{d }{4.93(11) \footnotemark[7]}\\
\hline \multicolumn{7}{c}{$\langle nsnp \,
^1\!P^o_1 ||D||nsmd\, {}^1\!D_2\rangle$}\\
 L-gauge & 4.43 & 4.62 & & 1.16 & 1.75 & 1.92
\\
  V-gauge      & 4.47 & 4.59 %&\mbox{---}&\mbox{---}&\mbox{---}&\mbox{---}
\\
%& 1.62 & 12.9 & 0.26 \\
 Final value       & \multicolumn{2}{d }{4.62(5)}
                    & \multicolumn{2}{d }{1.2(3)}
                    & \multicolumn{2}{d }{1.9(4)}
\\
 Experiment & \multicolumn{2}{d }{}
                & \multicolumn{2}{d }{}
                & \multicolumn{2}{d }{1.24(18)\footnotemark[8]}  \\
%               & \multicolumn{2}{c }{}
%               & \multicolumn{2}{c }{}
%               & \multicolumn{2}{c}{} \\
\hline \hline
\end{tabular}
%\end{ruledtabular}
\footnotemark[1]{Ref.~\cite{L80};}~%
\footnotemark[2]{Ref.~\cite{Zin};}~%
\footnotemark[3]{Ref.~\cite{KM80};}
\footnotemark[4]{Ref.~\cite{Lund};}
\footnotemark[5]{Ref.~\cite{PRT};}
\footnotemark[6]{Ref.~\cite{Smith};}
\footnotemark[7]{Ref.~\cite{Hans};}
\footnotemark[8]{Ref.~\cite{Hunter}.}
\end{table}

%####################################################################
\begin{table}
\caption{ Reduced electric-dipole matrix elements for {\em
intercombination} transitions. $n$ is the principal quantum number
of the first valence $s$ and $p$ shells and $m$ corresponds to the
first valence $d$ shell; $n=3$ for Mg, 4 for Ca,  and 5 for Sr;
$m=3$ for Mg and Ca, and 4 for Sr. All values are in a.u.}
\label{Tab_E1_inter}
%\begin{ruledtabular}
\begin{tabular}{ldddddd}
\hline \hline
&\multicolumn{2}{c}{Mg} & \multicolumn{2}{c}{Ca} &
\multicolumn{2}{c}{Sr}
\\
&\multicolumn{1}{c}{CI} & \multicolumn{1}{c}{CI+MBPT} &
\multicolumn{1}{c}{CI} & \multicolumn{1}{c}{CI+MBPT} &
\multicolumn{1}{c}{CI} & \multicolumn{1}{c}{CI+MBPT}
\\
%\hline

\hline
\multicolumn{7}{c}{$ \langle nsnp\,^3\!P^o_1||D||ns^2\,^1\!S_0\rangle$}\\
  L-gauge      & 0.0055 & 0.0064 & 0.027  & 0.034  & 0.12  & 0.16  \\
  V-gauge      & 0.0062 & 0.0062 & 0.030  & 0.032  & 0.13  & 0.17  \\
 Final value       & \multicolumn{2}{d }{0.0064(7)}
                    & \multicolumn{2}{d }{0.034(4)}
                    & \multicolumn{2}{d }{0.160(15)} \\
 Experiment    & \multicolumn{2}{d }{0.0053(3) \footnotemark[1]}
                & \multicolumn{2}{d }{0.0357(4) \footnotemark[2]}
                & \multicolumn{2}{d }{0.1555(16)\footnotemark[3]}
\\             & \multicolumn{2}{d }{0.0056(4) \footnotemark[4]}
                & \multicolumn{2}{d }{0.0352(10)\footnotemark[5]}
                & \multicolumn{2}{d }{0.1510(18)\footnotemark[5]}
\\             & \multicolumn{2}{d }{0.0061(10)\footnotemark[6]}
                & \multicolumn{2}{d }{0.0357(16)\footnotemark[7]}
                & \multicolumn{2}{d }{0.1486(17)\footnotemark[8]} \\
\hline

\multicolumn{7}{c}{$\langle nsnp\,^1\!P^o_1|| D ||
ns(n+1)s\,^3\!S_1\rangle$}\\
  L-gauge      & 0.0088 & 0.0097 & 0.035 & 0.043 & 0.15 & 0.19 \\
  V-gauge      & 0.0089 & 0.0101 & 0.035 & 0.045 & 0.15 & 0.20 \\
 Final value       & \multicolumn{2}{d }{0.0097(10)}
                    & \multicolumn{2}{d }{0.043(5)}
                    & \multicolumn{2}{d }{0.19(2)} \\
\hline
\multicolumn{7}{c}{$\langle nsnp\,^3\!P^o_1||D|| nsmd\,^1\!D_2\rangle$}\\
  L-gauge      & 0.0052 & 0.0049 &          & 0.059 & 0.33 & 0.19 \\
  V-gauge      & 0.0050 & 0.0047 &          & 0.061 & 0.36 & 0.18 \\
 Final value       & \multicolumn{2}{d }{0.0049(5)}
                    & \multicolumn{2}{d }{0.059(6)}
                    & \multicolumn{2}{d }{0.19(2)} \\
\hline
\multicolumn{7}{c}{$\langle nsnp\,^3\!P^o_2||D|| nsmd\,^1\!D_2 \rangle$}\\
  L-gauge      & 0.0039 & 0.0031 &          & 0.028 & 0.15 & 0.10 \\
  V-gauge      & 0.0041 & 0.0032 &          & 0.024 & 0.16 & 0.06 \\
 Final value       & \multicolumn{2}{d }{0.0031(4)}
                    & \multicolumn{2}{d }{0.028(3)}
                    & \multicolumn{2}{d }{0.10(2)} \\
\hline \hline
\end{tabular}
%\end{ruledtabular}
%\footnotemark[8]{}
%\footnotemark[9]{Ref.~\cite{Hunter};}
\footnotemark[1]{Ref.~\cite{God};}~%
\footnotemark[2]{Ref.~\cite{HusR};}~%
\footnotemark[3]{Ref.~\cite{HusS};}
\footnotemark[4]{Ref.~\cite{Kwong};}
\footnotemark[5]{Ref.~\cite{Droz};}
\footnotemark[6]{Ref.~\cite{Mitch};}
\footnotemark[7]{Ref.~\cite{Whit};}
\footnotemark[8]{Ref.~\cite{Kell}.}
\end{table}

%####################################################################
\bibliographystyle{apsrev}

\begin{thebibliography}{99}
\bibitem{Sap98}
J.~Sapirstein, Rev.\ Mod.\ Phys. \textbf{70}, 55 (1998) and
references therein.
\bibitem{SafJohDer99} see, for example,
M.S.~Safronova, W.R.~Johnson, and A.~Derevianko, Phys.\ Rev.\ A
{\bf 60}, 4476 (1999).
\bibitem{Zin}
G.~Zinner, T.~Binnewies, F.~Riehle, and E.~Tiemann,
Phys.~Rev.~Lett. {\bf 85}, 2292 (2000).
\bibitem{Seng}
K.~Sengstock, U.~Sterr, G.~Hennig, D.~Bettermann, J.H.~Muller, and
W.~Ertmer, Opt.~Commun. {\bf 103}, 73 (1993).
\bibitem{L80}
L.~Liljeby, A.~Lindgard, S.~Mannervik, E.~Veje, and B.~Jelencovic,
Phys.~Scr. {\bf 21}, 805 (1980);
\bibitem{FF}
P.~J\"{o}nsson, and C.~F. Fischer, J.~Phys.~B {\bf 30}, 5861
(1997).
\bibitem{JonFisGod99} P.~J\"{o}nsson, C.F.~Fischer, and
M.R.~Godefroid, J.\ Phys.\ B {\bf 32}, 1233 (1999).
\bibitem{VaeGodHan88} N. Vaeck, M. Godefroid, and J.E. Hansen,
Phys.\ Rev.\ A {\bf 38} 2830 (1988).
\bibitem{BFV}
T.~Brage, C.F.~Fischer, N.~Vaeck, M.~Godefroid, and A.~Hibbert,
Phys.~Scr. {\bf 48}, 533 (1993) (and references therein).
\bibitem{WGTG}
H.~G.~C.~Werij, C.~H.~Greene, C.~E.~Theodosiou and A.~Gallagher,
Phys.~Rev.~A {\bf 46}, 1248 (1992) (and references therein).
\bibitem{KM80}
F.~M.~Kelly and M.~S.~Mathur, Can.~J.~Phys. {\bf 58}, 1416 (1980).
\bibitem{Lund}
L.~Lundin, B.~Engman, J.~Hilke, and I.~Martinson, Phys.~Scr. {\bf
8}, 274 (1973).
\bibitem{PRT}
W.~H.~Parkinson, E.~M.~Reeves, and F.~S.~Tomkins, J.~Phys.~B {\bf
9}, 157 (1976).
\bibitem{Smith}
W.~W.~Smith and A.~Gallagher, Phys.~Rev.~A {\bf 145}, 26 (1966).
\bibitem{Hans}
W.~J.~Hansen, J.~Phys.~B {\bf 16}, 2309 (1983).
\bibitem{God}
A.~Godone and C.~Novero, Phys.~Rev.~A {\bf 45}, 1717 (1992).
\bibitem{HusR}
D.~Husain and G.~J.~Roberts, J.~Chem.~Soc.~Faraday~Trans.~2 {\bf
82}, 1921 (1986).
\bibitem{HusS}
D.~Husain and J.~Schifino, J.~Chem.~Soc.~Faraday~Trans.~2 {\bf
80}, 321 (1984).
\bibitem{Kwong}
H.~S.~Kwong, P.~L.~Smith, and W.~H.~Parkinson, Rhys.~Rev.A {\bf
25}, 2629 (1982).
\bibitem{Droz}
R.~Drozdowski, M.~Ignasiuk, J.~Kwela, and J.~Heldt, Z.~Phys.~D
{\bf 41}, 125 (1997).
\bibitem{Mitch}
C.~Mitchell, J.~Phys.~B {\bf 8}, 25 (1975).
\bibitem{Whit}
P.~G.~Whitkop and J.~R.~Wiesenfeld, Chem.~Phys.~Lett. {\bf 69},
457 (1980).
\bibitem{Kell}
J.~F.~Kelly, M.~Harris, and A.~Gallagher, Phys.~Rev.~A {\bf 37},
2354 (1988).
\bibitem{Hunter}
L.~R.~Hunter, W.~A.~Walker, and D.~S.~Weiss, Phys.~Rev.~Lett. {\bf
56}, 823 (1986).
%
\bibitem{PorKozRah00} S.G.Porsev, M.G.Kozlov, and Yu.G.Rakhlina,
Pis'ma Zh.\ Eksp.\ Theor.\ Fiz.\ {\bf 72}, 862 (2000)[ JETP Lett.,
{\bf 72}, 595 (2000)].
%
\bibitem{DFK}
V.~A.~Dzuba, V.~V.~Flambaum, and M.~G.~Kozlov,
Pis'ma~Zh.~Eksp.~Teor.~Fiz. {\bf 63}, 844 (1996) [JETP Lett. {\bf
63}, 882 (1996)]; Phys.~Rev.~A {\bf 54}, 3948 (1996).
\bibitem{DKPF}
V.~A.~Dzuba, M.~G.~Kozlov, S.~G.~Porsev, and V.~V.~Flambaum,
Zh.~Eksp.~Theor.~Fiz. {\bf 114}, 1636 (1998) [JETP {\bf 87}, 885
(1998)].
\bibitem{Din}
T.~P.~Dinneen, K.~R.~Vogel, E.~Arimondo, J.~L.Hall, and A.~Gallagher,
Phys.~Rev.~A {\bf 59}, 1216 (1999).
\bibitem{Kuro}
T.~Kurosu and F.~Shimizu, Jpn.~J.~Appl.~Phys., Part 2 {\bf 29},
L2127 (1990).
\bibitem{Kat}
H.~Katori, T.~Ido, Y.~Isoya, and M.~Kuwata-Gonokami,
Phys.~Rev.~Lett. {\bf 82}, 1116 (1999).
\bibitem{Mg}
M.~Machholm, P.~S.~Julienne, and K.-A.~Suominen, Phys.~Rev.~A {\bf
59}, R4113 (1999);
\bibitem{DalDav66}
A.~Dalgarno and W.~D.~Davidson, Adv.~At.~Mol.~Phys. {\bf 2}, 1
(1966).
\bibitem{Opt}
M.~G.~Kozlov and S.~G.~Porsev, Opt. Spektrosk. {\bf 87}, 384
(1999). [Opt. Spectrosc. {\bf 87}, 352 (1999)].
\bibitem{Thoul} see, for example,
D.~J.~Thouless, {\it The Quantum Mechanics of Many-Body Systems},
Chapter IV (Academic, New-York, 1975).
\bibitem{PRK1}
S.~G.~Porsev, Yu.~G.~Rakhlina, and M.~G.~Kozlov, J.~Phys.~B {\bf
32}, 1113 (1999).
\bibitem{PRK2}
S.~G.~Porsev, Yu.~G.~Rakhlina, and M.~G.~Kozlov, Phys.~Rev.~A {\bf
60}, 2781 (1999).
\bibitem{sidenote} Although the calculations were {\em ab initio} relativistic,
for brevity we suppress total angular momentum $j$ in the
designations of orbitals.
\bibitem{Bogdan}
P.~Bogdanovich and G.~\v{Z}ukauskas, Sov.~Phys.~Collection, {\bf
23}, 13 (1983).
\bibitem{Moore}
C.~E.~Moore, {\it Atomic Energy Levels}, Natl.~Bur.~Stand. (U.S.)
Circ. No. 467 (U.S., Washington, 1958).
\bibitem{Tiemann} E. Tiemann, private communication.
\bibitem{Bausch}
C.~W.~Bauschlicher Jr, S.~R.~Langhoff, and H.~Partridge,
J.~Phys.~B {\bf 18}, 1523 (1985).

\end{thebibliography}
%####################################################################

%####################################################################

\end{document}